\documentclass[journal]{IEEEtran}
\usepackage{cite}
\usepackage{times,amsmath,amssymb,psfrag,stfloats}
\usepackage{tabularx}
\usepackage{graphicx}
\usepackage{epsfig}
\usepackage{epstopdf}
\graphicspath{{./pics/}} \DeclareGraphicsExtensions{.pdf,.jpg,.png,.eps}
\usepackage{color}
\usepackage{algorithmic}
\usepackage{algorithm}
\usepackage[bookmarks=false]{hyperref}
\usepackage{multirow}
\usepackage{amssymb}
\usepackage{caption}
\usepackage{flushend}

\newcommand{\mo}{\color{black}}
\newcommand{\yc}{\color{black}}

\renewcommand{\algorithmicrequire}{\textbf{Initialization:}}
\renewcommand{\algorithmicensure}{\textbf{Iteration:}}

\renewcommand{\baselinestretch}{1}

\begin{document}
\title{Inference of Tampered Smart Meters with Validations from Feeder-Level Power Injections }

\author{\IEEEauthorblockN{Yachen Tang\IEEEauthorrefmark{1},
Chee-Wooi Ten\IEEEauthorrefmark{2}, and
Kevin P. Schneider\IEEEauthorrefmark{3}}\\
\IEEEauthorblockA{\IEEEauthorrefmark{1}Global Energy Interconnection Research Institute North America, San Jose, CA 95134 USA}\\
\IEEEauthorblockA{\IEEEauthorrefmark{2}Michigan Technologial University, Houghton, MI 49931 USA}\\
\IEEEauthorblockA{\IEEEauthorrefmark{3}Pacific Northwest National Laboratory, Richland, WA 99352 USA}
\thanks{This work is supported by US National Science Foundation (NSF) projects 1541000 and State Grid Corporation Technology Project 5455HJ180018.}}

\maketitle

\begin{abstract}
Tampering of metering infrastructure of an electrical distribution system can significantly cause customers' billing discrepancy. The large-scale deployment of smart meters may potentially be tampered by malware by propagating their agents to other IP-based meters. Such a possibility is to pivot through the physical perimeters of a smart meter. While this framework may help utilities to accurately energy consumption information on the regular basis, it is challenging to identify malicious meters when there is a large number of users that are exploited to vulnerability and kWh information being altered. This paper presents a reconfiguration switching scheme based on graph theory incorporating the concept of distributed generators to accelerate the anomaly localization process within an electrical distribution network. First, a data form transformation from a visualized grid topology to a graph with vertices and edges is presented. {\yc A conversion from the graph representation to machine recognized matrix representation is then performed. The connection of the grid topology is illustrated as an adjacency or incidence matrix for the following analysis. A switching procedure to change elements in the topological matrix is used to detect and localize the tampered node or cluster. The procedure has to meet the electrical and the temporary closed-loop operational constraints.} The customer-level anomaly detection is then performed in accordance with probability derived from smart meter anomalies.
\end{abstract}

\IEEEpeerreviewmaketitle

\section{Introduction}
\IEEEPARstart{E}{NERGY} thieves have been known as the major pitfalls for developing countries as the illegal electricity users are not part of the payment to utilities \cite{td1}. Similarly, alteration of individual's electronic energy usage on a smart meter is a covert, fraudulent behavior. Such motivation is to avoid full payment over the course of a long period of time. While smart meters provide the technology to conveniently obtain kWh information to utility billing center, poor physical perimeters may enable individuals to tamper the smart meters. Collectively, such an attack can get more creative that may be propagated through the Intranet of all associated smart meter devices to be enumerated by pivoting through a compromised smart meter \cite{type, svm}. Massive tampering can occur within the home area network (HAN) and neighborhood area network (NAN). Although this may not affect overall reliability of a distribution network, it would implicate the collection of monthly payments from utility's customers, leading to significant lower revenue as part of the financial losses \cite{loss, conf21}. On the other extreme, the discrepancy may trigger complaints by other customers who do not use usage much but amplified by the malware agents to significantly higher usage.

{\yc The metering data has the risk of being tampered during the process of measurement, collection, storage, or transmission while the malware or the worm propagation is usually embedded in the data string of the plausible legitimate metering data.} In time-critical communications, {\yc the data collection devices usually have access for security, the inquire of data is restricted. Therefore, it is hard to detect the existed malware in these devices. Mike Davis demonstrated the speed of the malware can break a smart meter's safeguard and worms propagation through the loophole in 2009 at Blackhat \cite{td2} while the Stuxnet worm attacked the metering units in the power system in 2010 \cite{td3,td4}.} Different than conventional malicious attacks, several smart meters might share the {\yc same identity number or a mechanical unit may generate one or more additional identity number to cause Distributed Denial-of-Service (DDoS) attack.} DDoS only affects the unavailability of smart meters. However, this attack can be utilized to reduce the share of resources in the topology and give attackers more information to perform other attacks \cite{td6,td7}. A McAfee report warned that {\mo an attacker could exploit smart meters easily} and takes control of the whole system \cite{td8}. Since the system specifications, diverse network protocols, and operational constraints in AMI, the existing defense methods against malicious attacks cannot be applied directly. {\mo The main motivation of this work is not to provide a defend method to improve the cybersecurity of the grid network. It proposed a switching operation scheme to detect and localize the occurred attacks or tampering behaviors within a distribution system.}

An assumption is that the AMI and SCADA systems have been fully deployed in the test area, however,the SCADA was developed using a comprehensive security policy to protect the system from the cyber attacks or any other threatens, while the validation for smart meters remains in the early stage. The main contribution of this paper is to {\yc combine the tamper detection method with the topology reconstruction switching procedure to locate} subfeeder or clusters with malicious meters in a distribution system. The {\mo distributed generators} are considered in this study to reduce the switching combinations. The rest of this paper is structured as follows: Related works are presented in Section II. The switching strategies are formularized in Section III. A case study with a virtual distribution network is discussed in Section IV. The conclusion and the future work are presented in Section V.

\section{Related Work}
Researchers to prevent anomalous electricity usages have done many existing methods. {\yc Artificial intelligence based machine learning and big data analysis \cite {c4} have been widely used for intelligent data analysis and data mining to identify normal consumption patterns, thus an obvious deviation between the normal pattern and the measuring data can be considered as an anomaly.} Authors in \cite{c8} using machine learning techniques on the past data to built models for detecting anomalous meter readings. A distributed intelligent framework using benford's law and stackelberg game to detect electricity theft was proposed in \cite{8088640}. {\yc A technique can identify diverse forms and caused deviations from tampered activities by using artificial neural networks\cite{td9} and Kalman filter\cite{c2}. The abnormal reports from the load areas with high tampered probabilities can be trained in support vector machine (SVM) \cite{svm} for classification.} Another main strategy of detecting malicious meters is based on the real-time comparison. In \cite{c1}, {\yc the (feeder) remote terminal unit (RTU/FRTU) provides a strategy to narrow down the search area for energy theft in a smart community.} Also, \cite{type} {\yc compares the real values with the predicted results to infer the possible tampered activities.} {\yc The methods to distinguish the legitimate data from metering devices and the malware-carrying data string and using a disassembler to assist mathematical analysis are described} in \cite{td11}. Different than data based detection approaches, \cite{td13} provides the attack and intrusion detection mechanism for a smart grid neighborhood area network (NAN). In addition, some strategies based on graph theory have been developed. In \cite{td12}, a novel inspection algorithm identifies each meter with a unique binary-coded number to locate the unique malicious meter is proposed. The conversion between the power grid and binary tree or spanning tree is adopted in \cite{svm}. {\yc Sensors can also be used to detect anomalies in customer levels and even in networks and communication levels. However, the deploy and maintenance fee of a complete sensor system could be expensive and false alarms may occur in the inspection process}\cite{c6}.

The reconfiguration switching scheme in the distribution system is mainly applied for the emergency operation and optimization process \cite{7294716}. The main purpose to exploit this problem is to achieve optimization. In \cite{c11}, {\yc the authors deal with the problem existed in the radial topology of the distribution network with minimum losses while \cite{c13, c18} demonstrated a strategy to optimize the cost for equipment and devices installation and maintenance via changing the states of switching devices. In \cite{c15} introduces methods to restore maximum load with minimum steps of switching procedures}.

\begin{figure*}[h!bt]
\centerline{\includegraphics[scale = 0.93] {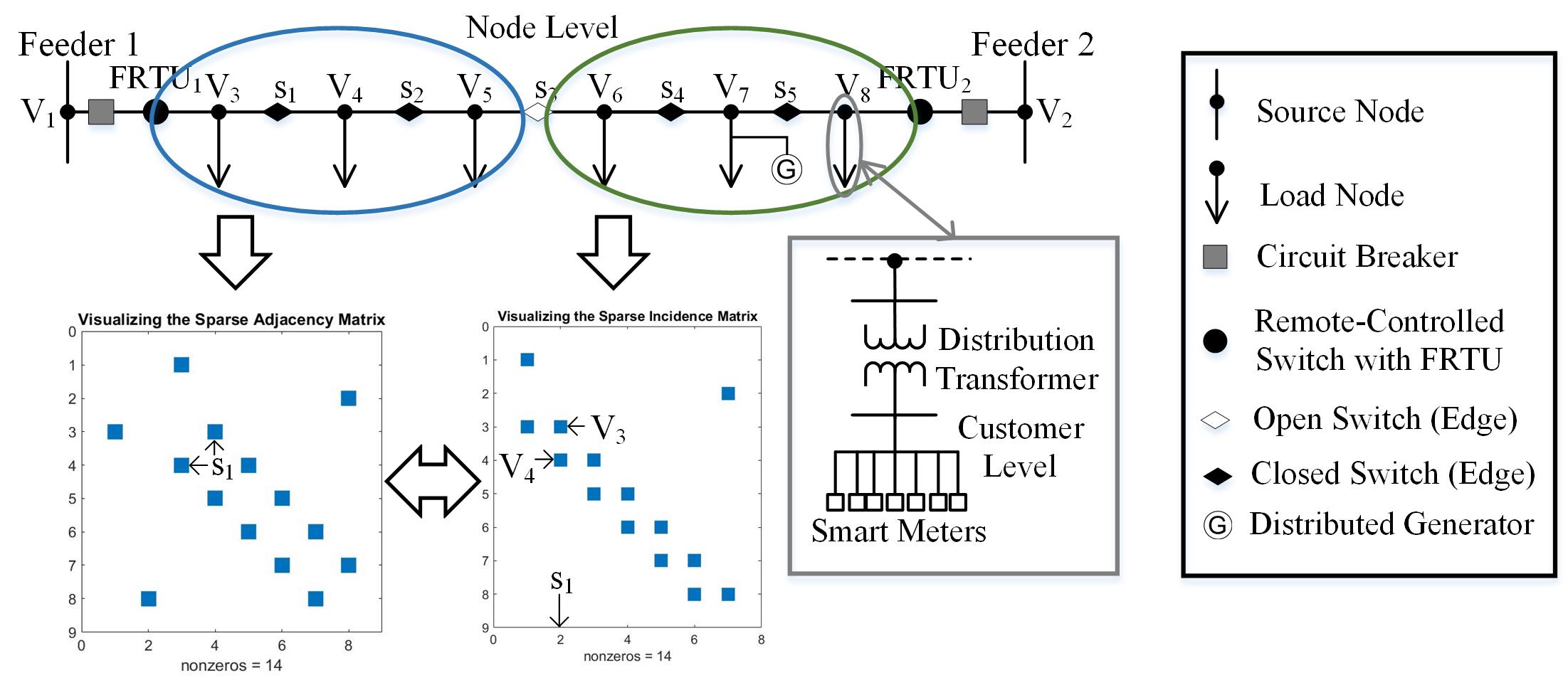}}
\caption{Sparse visualization of the adjacency and incidence matrices for the provided virtual topology.}\label{ftd1}
\end{figure*}

\section{Switching Strategies for Tampered Node Localization}
The profile-based anomaly detection to detect the irregularity of potentially tampered data by comparing the summation of all smart meters results with the metering result shown in the FRTU at the feeder head \cite{8088673}. {\yc If the offset is quite different (much larger or smaller) than the pre-defined threshold}, the reconfiguration switching scheme is utilized to localize the subfeeder, microgrid \cite{7587110}, cluster, or the distribution node with tampered load(s) or malicious meter(s). Since the operational complexity and the cost during the topological reconfiguration, the proposed method is  performed once the discrepancy is larger than 20\% of the normal power consumption (which will cause massive tampering). All electrical utilities keep the power on during the switching procedures while meeting the closed-loop electrical and operational constraints in the distribution grid \cite{6508558,8086023}.

\subsection{Convert the Distribution Network To a Graph}
{\yc A distribution grid is a radial topology and constructed by different feeders and microgrids. The adjacent feeders are connected with each other through tie switches (normally open and can be closed under emergency situation). The distribution grid can be represented by a graph $G = (V, E)$, which only contains nodes and edges. In $G$, all distributed power resources and loads are denoted as nodes/vertices $V$ while all nodes are connected by feeder lines with openable tie switches, which are referred to as edges $E$. The graph is stored in matrix form (adjacency and/or incidence matrix) to straightforward demonstrate the connection relationships between nodes and edges \cite{textbook}.}

\begin{algorithm}[h!tb]
\caption{{\mo Tampered Node} Localization Algorithm}
\label{algorithm2}
\begin{algorithmic}[1]
\renewcommand{\algorithmicrequire}{ \textbf{Input:}}{}
\REQUIRE ~~\\
$M_i, V_r, V_s, I_f$
\STATE Isolate all {\mo subfeeders or clusters with distributed generators}.
\IF {$I_f$ shows no {\mo tampered} load(s) in this system.\\}
\STATE $V_f = V_g$.
\STATE Restore the connection of each {\mo subfeeder or cluster} sequentially and check $I_f$.
\ELSE
\renewcommand{\algorithmicensure}{ \textbf{Iteration Fuction:}}
\ENSURE $F_f\ (M_i, V_r, V_s, I_f)$\\
\STATE $V_r(I_f) = 0$;
\STATE $\triangleright$ $V_r$ keeps updating;
\STATE $\triangleright$ {\yc Convert the topological incidence matrix to adjacency form and update the current states of all switches;}
\STATE $M_i \leftarrow M_i \cdot V_r^T$.
\STATE $M_a \leftarrow M_i \cdot M_i^T$.
\STATE Replace all diagonal elements in $M_a$ with 0.
\STATE $\triangleright$ Initialization;
\STATE $V_f \leftarrow V_s$.
\STATE $\widehat{V}_f \leftarrow V_f \cdot 0$.
\STATE $\triangleright$ Check if the status is same as the previous iteration.
\WHILE {$V_f - \widehat{V}_f \neq 0$}
    \STATE $\widehat{V}_f \leftarrow V_f$.
    \STATE $V_f \leftarrow \widehat{V}_f \cdot M_a + \widehat{V}_f$.
    \ENDWHILE
\STATE Replace all nonzero elements in $V_f$ with 1.
\RETURN $V_f$.
\STATE $\triangleright$ Localize the {\mo tampered node}.
\STATE Find all 0 elements in $V_f$.
\WHILE {More than one tampered nodes in one feeder, change $I_f$ to generate a new $V_r$.}
\STATE Repeat $F_f\ (M_a, V_r, V_s, I_f)$.
\ENDWHILE
\ENDIF
\end{algorithmic}
\end{algorithm}

An example of a distribution network with two feeders and the corresponding adjacency and incidence matrices are shown in Fig. \ref{ftd1}. The {\yc sparse} visualization of the adjacency and incidence matrices for the {\yc provided virtual} topology is also illustrated in Fig. \ref{ftd1}. The $FRTU_1$ is responsible to monitor the consumption states of all customers in the blue area while the $FRTU_2$ metering the green area. The detailed structure of the secondary network (customer level) for each load is shown in the gray block. In other words, the metering results in FRTU should be equal or slightly different than the summation of all smart meters' results under the supervised zone. It should be noted that the order of vertices and edges may change the form of matrices but can not change the topology of the graph. In this example, the order of vertices following the numbering sequence in this figure while the $S_1$ to $S_5$ represent edges 2 to 7. The $FRTU_1$ and $FRTU_2$ represent edge 1 and 7, respectively. Two vertices $V_3$ and $V_4$ and their connection $S_1$ (edge 2) is displayed on the visualized sparse matrices. The nonzero elements of these two matrices represent the number of edges in the graph. Since the graph is undirected, each edge is counted twice. This example is used for the case study in this paper.

\subsection{Tampered Node Localization with Switching Strategies}

{\yc The Algorithm \ref{algorithm2} illustrates the iterative process of how the node with tampered load(s) is localized based on the distribution switching procedure. The detailed descriptions of variables are shown as below:}

\addcontentsline{toc}{section}{Nomenclature}
\begin{IEEEdescription}[\IEEEusemathlabelsep\IEEEsetlabelwidth{$V_1$}]
\makeatletter
\renewcommand{\subsection}{\@startsection{subsection}{2}{4mm}
  {-\baselineskip}{0.5\baselineskip}{\it\leftline}}
\makeatother
\item[$M_i$] {\yc Topological incidence matrix to represent the connection states between nodes: $i$ is denoted as a node/vertex and $j$ denoted as an edge. If node $i$ is connected by edge $j$ to other node(s), $M_i[i,j]$ = 1, otherwise, $M_i[i,j]$ = 0.}
\item[$M_a$] {\yc Topological adjacent matrix to represent the connection states between nodes: both $i$ and $j$ are denoted as nodes/vertices. If node $i$ is connected with node $j$ , $M_a[i,j]$ = 1, otherwise, $M_a[i,j]$ = 0.}
\item[$V_r$] {\yc A defined row vector with length $\left|E\right|$ to indicate the connection states (open or close) of switches (edges) in the distribution topology. If switch $i$ is open $V_r[i]$ = 0, otherwise, $V_r[i]$ = 1.}
\item[$V_s$] {\yc A defined row vector with length $\left|V\right|$ to indicate the locations of all distributed power resources in the distribution topology. If node $i$ connected with a power source, $V_s[i]$ = 1, otherwise, $V_s[i]$ = 0.}
\item[$I_f$] Index of RTU/FRTU that indicates there exists {\mo tampered} load(s).
\item[$V_g$] {\yc A defined row vector with length $\left|V\right|$ to indicate} the location of distributed generators. If vertex $i$ connected with a distributed generators, then $V_g[i]$ = 1, otherwise, $V_g[i]$ = 0.
\end{IEEEdescription}

The algorithm starts with balancing the benefits of detecting the fraud, only the detected power losses are more than 20\%, the process is performed. The algorithm is summarized as follows:

\begin{enumerate}
  \item Isolate all subfeeders or clusters with distributed generators, if $I_f$ shows no {\mo tampered} load(s), the tampered load(s) exists in the {\mo isolated subfeeders or clusters}.
  \item Restore the connection of each microgrid sequentially and check $I_f$ to identify which microgrid has the tampered load(s).
  \item If the tampered load(s) is not in the isolated area, perform the iteration function to localize the tampered node.
  \item {\yc Convert the topological incidence matrix to adjacency form and update the current states of all switches for detecting the possible tampered node;}
  \item Initialize $V_f$ with $V_s$ and set $\widehat{V}_f$ as the previous iteration result of $V_f$ and initialize $\widehat{V}_f$ with 0;
  \item Check the status of every node during each iteration by checking $V_f = \widehat{V}_f$;
  \item {\yc If values of all nodes in the row vector are unchanged in the following iterations, use 1 to replace all nonzero elements in $V_f$};
  \item {\yc If there are more than one tampered nodes within a single feeder, update $I_f$ to generate a new $V_r$ and then repeat $F_f\ (M_i, V_r, V_s, I_f)$};
  \item Return $V_f$;
  \item {\yc 0 element(s) in $V_f$ to represent the location(s) of tampered node(s)}.
\end{enumerate}

\subsection{Customer-Level Anomaly Detection}
In this study, we assume all customers have smart meters. {\yc Continue the profile-based tampering detection process in consumers' level by comparing the historical consumption records with the current period report after the localization of tampered node in the load (or lumped load) level.} The anomaly score and the attacking probability of smart meters will be considered at the beginning of the customer-level anomaly detection to improve the searching efficiency. The searching process will start from the distribution node covers the smart meters with the highest anomaly index. {\yc The anomaly score of the smart meter can be estimated by a statistical approach. The ratio between the recorded number of anomalous readings and the total number of readings over a test period represents the score.} The anomaly score can be represented as (\ref{ano}):

\begin{equation}
\label{ano}
S_{\text{a}} = \frac{N_{\text{a},t}}{N_{\text{r},t}}
\end{equation}

\noindent where $N_{\text{a},t}$ is number of anomalous readings of a smart meter in a time duration $t$ and $N_{\text{r},t}$ represents the total number of metering results.

Assume the alarm system in each smart meter is sensitive, the attacking probability indicates the probability that an alarm occurs during a time duration of $t$, which can be estimated as (\ref{palm}):

\begin{equation}
\label{palm}
P_{\text{a}} = 1 - \prod (1-Q_{m}), ~m = 1, 2, 3, \cdots.
\end{equation}

\noindent where $Q_{m}$ represents the occurrence probability of type $m$ alarm, which can be estimated from the statistics and observations of archived data. There might be more than one tampering on one node.

\section{A Case Study}
{\yc A test case is utilized to validate the proposed strategy for detecting tampered meters based on the switching procedure. The schemes of changing connection status of switches in the distribution topology have to keep the whole system power on and meet the electrical and operational constraints.}

\begin{figure*}[h!bt]
\centerline{\includegraphics[scale = 1] {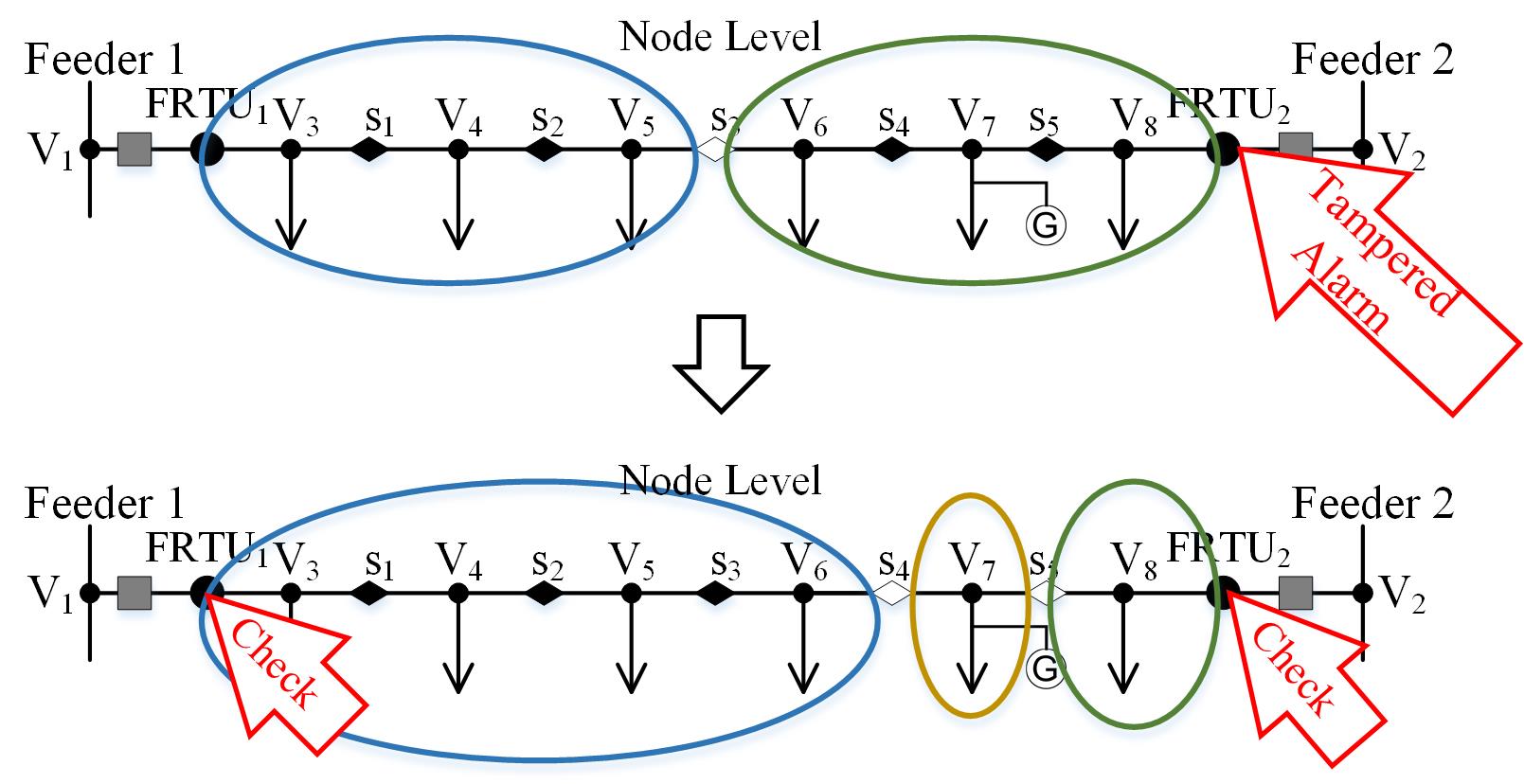}}
\captionsetup{justification=centering}
\caption{Switching procedures to localize the {\mo tampered} load node in the example topology.}\label{ftd2}
\end{figure*}

\subsection{Test Case Description}
Fig. \ref{ftd2} is the topology of the distribution test network. There are two feeders and six load nodes. The two remote-controlled switches at each feeder head are deployed with a FRTU. An alarm system based on the profile-based tampering detection is also installed in FRTU. The load node with the distributed generator can generate the microgrids. A random generated massive tampering alarm is sent from $FRTU_2$.

The corresponding input incidence matrix can be gathered from the previous section. The input row vectors of the first scenario in Fig. \ref{ftd2} are as $V_s = [1\ 0\ 0\ 0\ 0\ 0\ 0\ 1]$ since $V_1$ and $V_2$ are two power sources; $V_r = [1\ 1\ 1\ 0\ 1\ 1\ 1]$ due to the normally open tie switch $S_3$; $I_f = 7$ for the random fraud alarm and $V_g= 7$ because of the distributed generator. In the second scenario, the $V_s$, $I_f$, and $V_g$ remain unchanged while $V_r = [1\ 1\ 1\ 1\ 0\ 0\ 1]$ since the switching procedure.

\subsection{Switching Procedures}
Since the topology of the test system keeps unchanged, the input incidence matrix has no changes. The difference between each iteration is the updated connecting status $V_r$.

\subsubsection{Step 1} Under the operational constraints, connect the tie switch between these two feeders in order to keep the test system power on during this procedure.

\subsubsection{Step 2} Isolate the load node $V_6$ since it connected with a distributed generator which can supply the power for this microgrid.

\subsubsection{Step 3} Perform the profile-based tampering detection at $FRTU_2$, if $FRTU_2$ shows normally, the {\mo tampered loads} are not in $V_7$.

\subsubsection{Step 4} Perform the profile-based tampering detection at $FRTU_1$, if $FRTU_1$ shows normally, the {\mo tampered loads} are in $V_6$; If not, the {\mo tampered loads} are in $V_5$.

In this work, a simple virtual distribution network is applied, {\yc since the key point of the proposed method is to find out the best sequence of the switching procedure and change the states of openable edges, the different sequence will cause different searching time: the time could be exponential or only spend a linear time} in a real distribution network. This is the preliminary work that intends to prove the feasibility of the proposed idea.

\section{Conclusion and Future Work}
In this paper, a new application is presented to infer  {\mo tampered node}  within distribution AMI network using graph-theoretic framework. The data-based {\mo tampering detection} method is utilized to help indicate the existing of {\mo attacks or tampered activities} in the system. The consideration of {\mo subfeeders, clusters, or microgrids with distributed generators} will reduce the number of combinations during the switching procedures that can accelerate the localization of the {\mo spot with energy theft}. In the process of localizing the tampered load node, the distribution network is converted to a graph only contains vertices and edges. The graph is demonstrated as incidence or adjacency matrix for the analysis. A localization of {\mo tampered nodes}  {\mo within a network} is introduced.

Validations can be further extended in the future to study a large-scale distribution framework using the proposed algorithm. The consideration of electrical and operational constraints, corresponding combination strategy, as well as a heuristic search algorithm will be incorporated as part of the future work. {\yc The real topology of a distribution system can be drawn based on a real map in the geographic information system (GIS).} An agent-based model {\mo which can be integrated with multiple modules into a single simulation environment such as the GridLAB-D \cite{gridlabd}} to generate random {\mo tampered nodes} and perform the power flow to meet the constraint requirements in the provided network will be conducted.

\bibliographystyle{IEEEtran}
\bibliography{IEEEabrv,RefDatabase}

\end{document}